\documentstyle[psfig,epsf]{europhys}
\def\etal{{\hbox{{\tenit\ et al.\/}\tenrm :\ }}}

\def\And{{\rm and\ }}

\newif\ifboo \boofalse

\def\Review#1{\boofalse{\it #1},}

\def\Vol#1{\ifboo Vol. {\bf #1}\else{\bf #1}\fi}
\def\Year#1{\ifboo #1\else(#1)\fi}

\def\Page#1{\ifboo {\rm p. #1}\else{\rm #1}\fi}
\begin{document}
\euro{0}{0}{0}{0}
\Date{8 March 2000}
\shorttitle{I. Kanter \etal SECURE AND LINEAR CRYPTOSYSYTEMS}
\title{Secure and linear cryptosystems using error-correcting codes}
%\author{I. Kanter\inst{1}, E. Kanter\inst{1} \And L. Ein-Dor\inst{1}}
%\institute{
%     \inst{1} Department of Physics, Bar-Ilan University, 
%              Ramat-Gan 52900, Israel}
\author{I. Kanter, E. Kanter \And L. Ein-Dor}
\institute{
      Department of Physics, Bar-Ilan University, 
              Ramat-Gan 52900, Israel}
%
%\rec{8 March 2000}{}
%
\maketitle
\vspace{-0.80cm}

\begin{abstract}
A public-key cryptosystem, digital signature and authentication 
procedures based on a Gallager-type parity-check error-correcting code 
are presented.  The complexity of the encryption and the decryption
processes scale linearly with the size of the plaintext Alice sends to Bob.
The public-key is pre-corrupted by Bob, whereas a private-noise
added by Alice to  a given fraction  of the ciphertext of 
each encrypted plaintext  serves to increase the secure channel and 
is the cornerstone for digital signatures and  authentication. 
Various scenarios are discussed including the possible actions of 
the opponent Oscar as an eavesdropper or as a disruptor.
\end{abstract}
\vspace{-0.8cm}
The goal of cryptography is to enable two people, usually referred to  as 
Alice and Bob, to communicate over an insecure channel in such a way that 
the  opponent Oscar cannot understand and decrypt the transmitted 
message.\cite{crypto}
A block message is called a plaintext, and a long message is a sequence
of plaintexts.  
In a general scenario, the plaintext is encrypted by Alice 
through the key $E_k$ and 
the result, ciphertext, is sent over the channel. A third party, 
eavesdropping on the channel, cannot determine what the plaintext was. 
However, Bob, who knows the encryption key, can decrypt the ciphertext using 
the key $D_k$ and recover the plaintext.

In a private-key system, the keys $E_k$ and $D_k$ are known only to Alice and 
Bob, and it obviously increases the security of the  channel. 
However, a private-key 
system requires communication between Alice and Bob prior to the 
transmission of any plaintext. This prerequisite
makes the private-key communication
impractical in modern communication, especially in such areas as 
electronic commerce and Internet-based communication.
The goal of public-key systems is to devise a cryptosystem where it is 
computationally infeasible to determine $D_k$ given $E_k$, and hence
the encryption rule $E_k$ can be made public.

The secure channel and the efficiency of a public-key cryptosystem
depends on many parameters, among them: (a) the complexity to determine 
$D_k$ given $E_k$; (b) the complexity of the encryption/decryption processes;
(c) the length of the ciphertext and the public-key in comparison to the 
length of the plaintext. 

The commonly used RSA cryptosystem\cite{RSA78} is based  on the difficulty 
of factorizing large integers.
Its main drawback is that 
the complexity of the encryption/decryption processes is of $O(N^2)/O(N^3)$, 
where $N$ is the length of the plaintext (see figure 1). For small
$N$ these complexities are also small, 
but then the complexity to determine $D_k$ given $E_k$ may also be 
accessible  to Oscar. It was recently found that even $N=512$ 
may be too small to ensure a secure channel.
Hence the complexity of the
encryption/decryption becomes the bottleneck of public-key cryptosystems
as well as for other tasks of the secure channel (digital
signature, authentication, etc.) based on such methods.

From the known cryptosystems it appears that there is a trade off between the 
secure channel and the complexity of the encryption/decryption processes.
In this work, we propose a new secure cryptosystem,
based on the preliminary bridge built 
between error-correcting codes and cryptosystems by McEliece\cite{McE}
with the following features and ingredients:
(a) The complexity of the encryption/decryption processes scale 
linearly with the size of the plaintext $N$. These complexities can be 
easily reduced even further under parallel dynamics.
(b) The method is based on boolean operations between
two sparse matrices, in contrast of
factorizing large integers in cryptosystems based on number theory.
(c) Our method  consists of many  stochastic ingredients; 
Bob adds noise to the public-key whereas Alice adds noise to the ciphertext. 
(d) The method is applicable as a public-key cryptosystem, as well as 
for  digital signatures and authentication. A digital signature 
is used to specify the person responsible for the message, and 
an authentication ensures the integrity of the plaintexts 
constructing the message.\cite{crypto}
It is a challenge to have a secure public-key cryptosystem operating 
with low complexity which can serve also for all the different 
tasks of the secure channel. 

Before  describing the details of our method, let us first categorize 
the possible capabilities of Oscar: (a) An eavesdropper: Oscar may try to
reveal the plaintext Alice is sending  to Bob from the transmitted ciphertext
and/or the digital signature.
(b) A distruptor: The message Alice is sending to Bob 
can be repeated, replaced or corrupted during 
transmission by Oscar. We may distinguish between a forged 
meaningful/meaningless  signed plaintext.
Note that the ability to forge many meaningless but 
legally signed messages
could cause a disastrous effect in the event of real-time procedures.
It may take some critical time for Bob to realize that 
legally signed messages are  forged messages rather than noisy ones.
Note that in cryptosystems such as RSA,\cite{RSA78}  it is easy  to forge 
a meaningless signed message or to repeat the transmission of the 
same message or previously legally signed messages to Bob.

Our cryptosystem is based on an 
error correcting method known as the Gallager method\cite{Gallager}
or its MN version.\cite{MacKay,kanter1,kanter2}
It comprises 
two sparse boolean matrices, $A$ and $B$,
of dimensionalities $M\!\times\! N$ and $M\! \times\! M$
respectively, and the rate $R \equiv N/M \le 1$. Note that all operations
are in $\mbox{(mod 2)}$.
The encryption public-key  is an $M \times N$ matrix
\begin{equation}
E_k = B^{-1}A \ \mbox{(mod 2)}
\end{equation}
\noindent and the encrypted ciphertext is 
\begin{equation}
C=E_k  \ s \ \mbox{(mod 2)}
\end{equation}
\noindent Before the transmission, Alice adds noise (bits flipped with 
probability $f$)
to $C$, such that the received ciphertext is 
\begin{equation}
r = C + n  \ \mbox{(mod 2)}
\end{equation}
\noindent where $n$ represents the noise.
The decryption key 
\begin{equation}
D_k = [A,B]
\end{equation}
\noindent is known only to Bob, who can find $s$ by 
multiplying $r$ with the matrix $B$ 
to obtain $z = B  \ ({E_k s}+{n}) = A s + B {n} \ $.
It requires solving the equation
\begin{equation} 
[A , B] \left[ \begin{array}{c}
{s '} \\ n '
\end{array} \right] =  z  \ , 
\end{equation}  
\noindent where $s'$ and $n'$ 
are the unknowns, but their statistics 
(for instance, unbiased message for $s '$, and flip rate $f$ 
for $ n '$) are known. 
This may be carried out 
using standard methods such as that of belief network 
decoding.\cite{MacKay}
In this method, representing a
special case of parity-check codes, each bit of the ciphertext $C$ 
is derived from the parity of a sum of certain plaintext's bits.
It has recently been shown, 
based on insight gained from the study of diluted spin systems\cite{Sompo}
that specific choices of 
cascading sub-matrices $A$ and $B$ can nearly saturate 
Shannon's bound.\cite{kanter1,kanter2}

With the lack of noise and invertible $E_k$ (otherwise  decryption  
cannot be terminated successfully with probability one), Oscar is able
to easily recover $s=E_k^{-1}r$.  
In order to make Oscar's task
more difficult, we follow the line of 
McEliese where noise is added to $C$, namely, bits are 
flipped ($0 \rightarrow 1$ or $ 1 \rightarrow 0$) with probability 
$f$.\cite{crypto}
In this event $E_k^{-1}r$ results in an approximated plaintext, $s_{app}$,
where a fraction of the bits are wrong.
For a given rate $R$ and large $N$, the maximal noise $f$ 
(for which the decryption could terminate successfully without error 
bits in the decrypted plaintext) is given
by the maximal channel capacity\cite{Cover}
$R_{c}=1-H_2(f)$ where 
$H_2(f)=-f \log_{2}(f)-(1-f) \log_{2}(1-f)$. 
Oscar's task to recover $s$ is difficult, since he has to decompose 
$E_k$ into the matrices $B^{-1}$ and $A$, which is known to 
be an NP-complete problem.\cite{GARY}

The main drawbacks of this cryptosystem are: 
(a) Strong finite size effects
which are  visible even for large $N=O(10^{4})$.
The decryption typically terminates with some 
percentages of error bits, which is catastrophic from a practical point of 
view. 
(b) For large $N$, the encryption evolves a product of a 
matrix $(M \times N)$ $E_k$ by the plaintext $s$, hence
its complexity is $O(N^2)$. 
Similarly, the complexity of each step of the decryption is  $O(N^2)$, 
eq. (5).
Clearly it is less than the
cubic complexity of the decryption in RSA.
However, if in practice one has to work with  $N=10^4$, 
the reduction in the complexity becomes a drawback.\cite{Saad}
Furthermore, the size of the public-key which has to be downloaded by Alice
diverges as   $O(N^2)$. 
%(d) Working reliably, if any, with  small $f$, Oscar can first find
%$s_{app}$ and then find the small number of wrong bits in an exhaustive search.

To overcome these drawbacks, which prevent the usefulness of error-correcting
codes as a cryptosystem,  we used the following three observations.
In error-correcting codes, the nature of the 
channel typically has a {\it statistical nature}; a probability for 
a bit to flip or a width of the Gaussian noise. Each transmitted bit has
the same probability to flip from a given distribution.\cite{Cover}
The first observation is that 
in a case where an error-correcting method serves as a cryptosystem, 
Bob can pre-corrupt the public-key
$E_k=B^{-1}A$ in the following sense. 
In a fraction $p_q$ of rows, part (all) of  the elements are flipped 
at random. 
The location of these $p_qM$ rows is known only to Bob.
Hence, a fraction $p_q$ of the ciphertext is corrupted with an average 
probability $1/2$. Since this pre-corruption of $E_k$ is common to 
all ciphertexts, we denote  it as a {\it quenched noise}, $n_q$. The  main 
purpose of the quenched noise is to make Oscar's task  of decomposing 
$E_k$ to $B^{-1}$ and $A$ more difficult.
Note that our cryptosystem works properly for some 
sub-classes of matrices $A$ and $B$, but for a random 
choice our cryptosystem fails with probability one. Hence the task of Oscar is
to find a decomposition which works properly as a cryptosystem.
The second observation: in addition to the pre-corruption process, 
Bob publicizes a {\it given} fraction, $p$, of the ciphertext where 
Alice's private-noise, $n_a$, can be added. 
This localized private-noise consists of 
a flip rate $f$  of {\it given} $pM$ bits of the ciphertext. 
The resulting ciphertext then comprises of 
frozen (non-flipped) bits,
randomly flipped bits and flipped bits with probability $f$.
The presence of frozen/flipped bits in the 
plaintext serves to increase the secure channel and to suppress
finite size effects. 
Similar to Shannon's bound,\cite{Cover} one can show that for a given rate $R$ 
the maximal fraction of flipped bits with probability $f$ is
\begin{equation}
p_c={1-p_q-R \over H_2(f)}
\end{equation}
\noindent We assume that  a fraction $p_q$ of the bits
are flipped with probability $1/2$,
however, $p_c$ might even be further improved for the following reason.
In an error-correction scenario 
only statistical properties of the plaintext and the flip rate 
are known, hence any decoded state obeying these statistical 
features is valid. 
In contrast, Bob knows 
the  manner in which $E_k$ was corrupted and 
hence the error in the $p_qM$ corrupted bits 
should be consistent with the decrypted plaintext. 

The most striking observation comes from
various simulations on different random constructions of the matrix $B$,
indicating the following rule.  As long as the {\it average} 
connectivity, number of non-zero elements per column,  
of $B$ is smaller than $2$, $B^{-1}$ is sparse. A random construction
means that the elements of each row are chosen at a random position, with no 
spatial  structure.
Since $A$ is a sparse matrix with random positions of the 
non-zero elements, it is clear that $E_k$ is also sparse.
Hence, for such  matrices $B$, 
the size of the public-key $E_k$ scales linearly with the size of the 
plaintext. Furthermore, the complexity of the decryption process also scales
linearly with the size of the plaintext,  as the number of iterations
is of $O(1)$ (see details below). A comparison with encryption/decryption
complexities of the RSA system\cite{RSA78} is presented 
in figure 1 for various sizes
of plaintexts.
A sparse public-key is a necessary requisite  for an efficient encryption
process of large plaintexts, which are of great practical importance.
For an average  connectivity greater than $2$, $B^{-1}$ is heavily dense,
and the number of non-zero elements in $E_k$ is around $MN/2$.

The sparseness of
$B$ for a connectivity less than $2$
can be supported by the following theoretical argument.
Assume that the matrix $B$ is constructed such that
$B_{ij}=\delta_{i,j}+\delta_{i+c,j}$ for $i \le \rho M$ 
and $B_{ij}=\delta_{i,j}$ for $i > \rho M$, where $\rho <1$ and
the average connectivity is  $1+\rho <2$.
This matrix can easily be inverted and one can show that for $c=O(1)$ 
$B^{-1}$ is dense (with a small prefactor), 
but for  $c=O(N)$, $B^{-1}$ is sparse.
In a random construction the typical distance between two non-zero
elements belonging to the same row is of $O(N)$. 
With respect to this property it is similar to $c=O(N)$. 
Furthermore, for connectivity below $2$ the generic graph represented by $B$
is below the percolation threshold.

We perform simulations on 
$R = 1/2$ and $256 \le N \le 2048$ with a few different classes  
of matrices $A$ and $B$ and here we report only limited results. 
Each parameterization of the matrices $A$ and $B$ 
was averaged over at least $10^5$ plaintexts and $50$ realizations.
The construction follows the spirit of the constructions for 
error-correcting codes of the Gaussian channel and $R=1/2$.\cite{kanter2}
The structure of the matrix $B$ is such that  
for $i \le \rho M$  there are two non-zero elements at random positions
where for $i > \rho M$ there is only one non-zero element. 
The matrix $A$ is constructed such that in the first 
$\rho' M$ rows there are two non-zero 
elements and in the remaining rows there are six non-zero 
elements chosen at random positions. In order to break the
inversion symmetry, where each row of $A$ consists of an even number of
non-zero elements, a small number of rows were changed from $2 \rightarrow 1$ 
and $7 \rightarrow 6$ non-zero
elements.
Note that the spatial separation between different rows of the matrices 
was done only for demonstration, and to increase the security of the 
channel one can mix their locations.
The performance depends on the success rate of the 
decryption as a function of ($N,p,p_q,f$), where the private-noise was
added to the first $pM$ bits of the ciphertext. 
%Many points in this four dimensional space were examined numerically. 
Let us present 
a few examples among many where the decryption terminates successfully over 
at least $10^5$ plaintexts in a finite fraction of the realizations:
(a) $\rho=1/2$, $\rho'=7/8$ and $(512,0.53,0-0.04,0.04)$,
(b) $\rho=\rho'=3/4$ and $(1024,0.53,0-0.04,0.075)$ and
(c) $\rho=3/4$, $\rho'=7/8$ and $(768,0.53,0-0.04,0.088)$.
These results indicate that the probability for a wrongly decrypted block
(plaintext) is $P_B < 10^5$.
In all the above-mentioned classes, the number of iterations
of the belief algorithm is typically $ \sim 10$ steps, 
where the complexity of each step of
the algorithm is of the order of the number of non-zero elements in 
matrices $A$ and $B$, $O(N)$. 
No long tail in the distribution of the convergence time 
was observed. Note that each belief iteration can be implemented 
in parallel such that the time complexity can be reduced by $O(1/N)$.
Results indicate that  finite size effects are dramatically suppressed by the 
frozen bits (in contrast to homogeneous noise), 
and can be improved  even further by increasing $N$. 
 
The location of frozen (non-flipped) bits of 
the ciphertext $(1-p)M$ is known also to Oscar. 
Hence, the secure channel forces
the number of frozen bits $(1-p)M <N$.  Otherwise 
Oscar may try to solve $E_k^{1-p}s_{1-p}=C_{1-p}$, where $1-p$ indicates the 
relevant part of the matrix/vector corresponding to the frozen bits. 
In such a case where  $E_k^{1-p}$ (of dimentionalities 
$(1-p)M \times N$) is invertible, Oscar can easily find the plaintext $s$. 

Let us now discuss  a possible attack on our cryptosystem. 
Oscar's goal is to find a partial public-key, $E_k^{part}$, 
obeying the following constraints: (a) The dimentionalies of $E_k^{part}$
is $M' \times N$ where $N \le M' \le M$. (b) The corresponding  $M'$ bits
of the ciphertext are the correct ones.
(c) $E_k^{part}$ is invertible. In such an event 
Oscar can easily find the plaintext $s$, and the question is, what is 
the probability of such an event? The number of frozen bits of the ciphertext
is $(1-p)M$ which was chosen to be less than $N$. Assuming these
$(1-p)M$ rows are linearly independent, Oscar has to guess additional
$N-(1-p)M=N(R+p-1)/R$ correct rows 
in order to construct a plausible invertible  $E_k^{part}$. 
The probability of such an event is $(1-f)^{N-M(1-p)}$ 
and it becomes negligible as we increase the size of our 
plaintext. Furthermore, in simulations we realized that the 
rank of the $(1-p)M$ correct rows is
is $\sim 0.9(1-p)M$.
Hence Oscar has to guess additional correct rows 
%of the public-key 
and the probability of such an event decreases even further.
Last but not least, how does Oscar know that he chose the correct rows?
A plausible answer to this question  is that any set of additional correct rows
results in the same  plaintext $s$. Hence, Oscar may repeat the 
above attack many times 
and deduce that the most probable outcome is the desired plaintext.
The first difficulty with this scenario is that 
$s$ has to appear many times as an outcome, since
we would like to distinguish between the signal, $s$,
and the noise of other possible outcomes.
Secondly, the most probable (exponentially dominated) 
$E_k^{part}$ consists of $(N-M(1-p))f>0$ wrong rows. 
It is true that  all these  $E_k^{part}$ do not necessarily result
in the same plaintext, but their distribution is still in question.

The secret information of Bob is the decomposition of the public-key 
$E_k$ into $B^{-1}$ and $A$ and the quenched noise ,$n_q$, used to corrupt 
the public-key. The secret information of Alice is the plaintext $s$ and 
the private noise, $n_a$.
The key point of our signature scheme is that after the decryption process 
terminates successfully Bob recovers not only the plaintext $s$ but also the 
private noise, $n_a$, added 
to the ciphertext. 
More precisely, on one hand side,
from the decryption of the plaintext $s$ Bob 
knows the corrupted ciphertext by using  
the corrupted public-key, $E_ks$. On the other hand, Bob has in his hand the 
received ciphertext, $E_ks+n_a$. From the difference 
between these two pieces of information Bob can easily find $n_a$. 
The ability of Bob to reveal  $n_a$ 
besides the plaintext is at the center of the discussion below. It
shows how
Alice can use the additional information to sign and to 
keep the integrity of the message. 

A simple signature 
is based on the following two ingredients:
(a) Alice constructs an additional  plaintext comprising of a linear 
combination of  $s$ and $n_a$, $X(s,n_a)$.
The new plaintext $X$ is encrypted by Alice to a new ciphertext, $t_1$,
using  $E_k$ and a new private noise $n_{a1}$. 
Alice transmits to Bob both ciphertexts, $t$ and $t_1$.
(b) For verification, Alice constructs by a known procedure
a verifiable vector, $V=V(s,n_a,n_{a1})$.
After the decryption of 
both ciphertexts Bob knows all the ingredients of $V$ and the verification 
can be carried out. Note that for a one-time signature scheme where
Oscar is functioning as an eavesdropper only, our channel is secure.
The usefulness of these signature schemes is twofold:
(a) The signature/verification procedure is very easy 
for Alice/Bob to implement with complexities of 
$O(N)$. (b) A plaintext repeated twice
has in each transmission a different signature due to the different 
private-noise. 
%Such a time dependent signature may  be used to identify the time 
%(or stamping) that Alice/Bob first encrypt/decrypt the message.
The main drawback of the above signature scheme is that 
Oscar can easily forge a legal plaintext. 
There are exponentially many
plaintexts $s$ and private-noise $n_a$ and $n_{a1}$ which give the same
verifiable vector $V$ 
%and each of them can be constructed with $O(N)$ steps. 

It appears that in order to have a secret personal signature Alice has 
(a) to send additional information besides the plaintext; and
(b) to use her own 
known signature scheme based on her  cryptosystem, similar to
using RSA for both the encryption and signature.\cite{crypto}
Surprisingly, we demonstrate below that Alice can construct a secure
signature without the transmission of any additional information
besides the encrypted plaintext. 

A simple scenario for an advanced  secure signature is one in which Alice 
first generates a vector $V$ of rank $N'<N$ using $s$ and $n_a$ following 
her public protocol.
Next, the number of $1$'s in $V$ is truncated to a 
fixed  number $K$  (or $\le K$) 
following Alice's public prescription. 
(For rare events where there are 
no $1$'s in $V$, Alice provides a special procedure).
The signature, $E_k^{part} V$ is left publicized 
by Alice, where $E_k^{part}$ stands for the relevant rows
corresponding to the rank $N'$.
Determining $V$ from the knowledge of $E_k$ and
the signature is known to be a NP-complete problem (pg. 280 in ref. (14)). 
Bob, who knows $s$ and $n_a$, can easily verify the signature. 
One can easily make the situation more complex when 
Alice creates a new cryptosystem in an on-line manner.
For instance, 
based on the public-key, $E_k$, 
Alice implements  some permutations among the rows as a function of the
detailed structure of $s$ (and/or $n_a$).
The permutation scheme is publicized  by Alice, which may be
fixed for all plaintexts/users or can be chosen time dependently. 
Let us denote the permute public-key by $E_k^P$, hence 
the signature of $s$  is $t_1=E_k^PV$, 
publicized by Alice 
but does not have to be transmitted over the channel.
Bob first decrypts $t$ and obtains $s$ and $n_a$. Next, Bob finds his
permuted public-key, $E_k^P$, using the public permutation prescription
chosen by Alice, and then easily verifies $t_1$ from $E_k^PV$.
Since the signature depends on $s$ and $n_a$ as well as  on 
$E_k$, the same plaintext 
transmitted to different addresses or at different times (different $n_a$) 
is characterized by different signatures.  
As an eavesdropping, Oscar 
does not know $s$, $n_a$ and also $E_k^P$. 
As a disruptor, Oscar 
may try to replace the ciphertext $t$ by $t'$ consisting of $s'$ and $n_a'$
such that the signature of $V=V(n_a')$ using 
the permuted public-key $E_k^P(s')$ has the same signature as that publicized 
by Alice. The lack of an independent permuted public-key as a function of the 
plaintext seems to make the work of a disruptor even harder.  
In principle Alice can corrupt  her signature $E_k^PV$ 
either by flipping some bits among a small fraction of the ciphertext
or by adding a noise consisting of $n_a$ and $s$.  
In such a scenario Bob has first to decrypt 
$s$ and $n_a$ and then to verify the 
decrypted signature with the public protocol of Alice. 
For a permuted public key the decryption follows
the permuted matrices $A_{per}=A$, and 
$B_{per}$ is nothing more than the same permutation as made for the 
rows of $E_k$, but  now for the columns of $B$. 

The aim of the authentication procedure is to keep the integrity of the message
constructed from a sequence of plaintexts, such that Oscar cannot
forge (add/delete) ciphertexts.
Using error correcting codes as a cryptosystem this goal can be 
achieved by using correlated noise for successive ciphertexts. Let 
us describe some possible scenarios.  The private-noise for the first
ciphertext is chosen as explained above, where   the noise 
for the next ciphertext is related to the previous one by some 
permutations.
The permutations may depend on (a) the private-noise of the previous 
ciphertext; (b) the previous plaintext; (c) both the previous ciphertext,
and the private-noise. 
One may think that the permuted noise  has to be bounded to  
the allowed regime by Bob  in order to ensure a successful 
decryption (which can be easily achieved).
However,  there is no necessity for such a restriction, 
since after the decryption of the first plaintext, 
the private-noise of the next ciphertext is uniquely determined 
and hence also the plaintext. 

The advantage of such an authentication scheme is that Bob has only   
to decrypt the first plaintext, whereas
the rest of the message is uniquely defined,
since the noise is known. 
On the other hand, Oscar knows the authentication scheme and may 
concentrate only on the decryption of the first ciphertext, or
alternatively on an
%, the decryption by Oscar of an 
intermediate  ciphertext (the easy one)
which reveals all  successive plaintexts.
In order to ensure the same security of (almost) all plaintexts, one can 
use accumulated permutations.  The private-noise for the current
ciphertext depends on all previous plaintexts/private-noise 
by a publicized procedure.

The tasks of our cryptosystem can be extended to other 
functions of the secure channel, such as an undeniable signature.
The private-noise is added 
out of the allowed range such that the decryption cannot terminate 
successfully without Alice partially revealing her noise.

\vspace{-0.20cm}
Alice has to keep as public information all 
previous signatures. The list of the signatures  
may load Alice's  resources, and furthermore it
may  take a long time for Bob to find the appropriate 
signature among many.
This drawback  can be alleviated by removing 
the signature into an archive
after verification by Bob.

\begin{figure}
\centerline{\psfig{figure=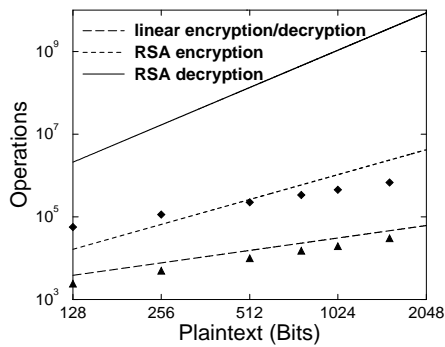,width=5.8cm}}
%\vbox to 1cm{\vfill\centerline{\fbox{1a.eps}}\vfill}
%
\caption{
Requested number of operations
to encrypt/decrypt a ciphertext  vs. the length of the plaintext in bits.
The encryption/decryption complexity  for the RSA cryptosystem is 
$O(N^2)/O(N^3)$ (solid/dashed lines)  and the prefactor is 
normalized to unity.
The averaged number of operations  obtained in  
simulations for $R=1/2$, $\rho=1/2$, $\rho'=3/4$ and 
$(N,p=0.53,p_q=0-0.04,f=0.045)$ are presented 
for the encryption/decryption (triangle/diamond) processes, where
error bars are less than the size of the symbols. 
As a guideline, a linear curve with a prefactor equals to $30$ is presented
(long-dashed line). Note that in the encryption, the complexity consists of
boolean operations (eqs 1-3), where the complexity of the decryption 
measures multiplications of real numbers, (eq. 5).
}
\label{fig1}
\end{figure}

In conclusion, let us briefly discuss a few of the advantages 
of our cryptosystem
over methods based on  numbers theory, such as an RSA cryptosystem. First,
the matrix operations/belief network decoding in the decryption/encryption
process can be carried out and implemented in parallel.
Secondly, a one-time success by Oscar to reveal a plaintext
does not automatically help or ensure the recovery of other 
plaintexts that  Alice sent to the same Bob. 
Finally, in the RSA method, for instance, Oscar's task requires
a check of many possible trails, where each trail  can be 
examined by the same algorithm. Hence, the task of Oscar can be easily
split among many resources. In contrast, our cryptosystem
is based on many  stochastic ingredients  with 
time dependent features of Alice and Bob. Hence the 
strategy of Oscar may need to vary between different messages and
users of the channel. Even for a given channel,   
the challenge for future research is to find 
how to parallelize and to simplify
the task of Oscar.

I. K. acknowledges fruitful discussions and 
comments by  W. Kinzel. The content of this work is patent pending.

\end{document}